# Rapid micro fluorescence *in situ* hybridization in tissue sections

D. Huber and G.V. Kaigala*

IBM Research – Zürich, Säumerstrasse 4, 8803 Rüschlikon, Switzerland

*corresponding author: gov@zurich.ibm.com

## Abstract

This paper describes a micro fluorescence *in situ* hybridization (µFISH)-based rapid detection of cytogenetic biomarkers on formalin-fixed paraffin embedded (FFPE) tissue sections. We demonstrated this method in the context of detecting human epidermal growth factor 2 (HER2) in breast tissue sections. This method uses a non-contact microfluidic scanning probe (MFP), which localizes FISH probes at the micrometer length-scale to selected cells of the tissue section. The scanning ability of the MFP allows for a versatile implementation of FISH on tissue sections. We demonstrated the use of oligonucleotide FISH probes in ethylene carbonate-based buffer enabling rapid hybridization within < 1 min for chromosome enumeration and 10-15 min for assessment of the HER2 status in FFPE sections. We further demonstrated recycling of FISH probes for multiple sequential tests using a defined volume of probes by forming hierarchical hydrodynamic flow confinements. This microscale method is compatible with the standard FISH protocols and with the Instant Quality (IQ) FISH assay, reduces the FISH probe consumption ~100-fold and the hybridization time 4-fold, resulting in an assay turnaround time of < 3 h. We believe rapid µFISH has the potential of being used in pathology workflows as a standalone method or in combination with other molecular methods for diagnostic and prognostic analysis of FFPE sections.

## I. INTRODUCTION

The most common cancer in females is breast cancer with 464'000 new cases alone in 2012, of which 131'000 cases were lethal[1]. While the breast cancer incidence rate has been increasing, the death rate has decreased by 40% since the 90ies[2]. This is likely due to early detection of the solid tumor linked to efforts in public health to improve awareness among women (at risk) and routine screening. Further, the detection of breast cancer related biomarkers through genomic and transcriptomic analysis allows reclassification of the tumors into molecular subtypes[3,4] such as Luminal A (ER+, PR+, HER2-), Luminal B (low p53 mutation, ER+, PR+, HER2++), Basal-like (Triple-negative) or HER2++ (Fig. 1a). Associated subtype-specific targeted therapies have led to more effective and successful treatment of patients and increased the survival rates[3,4].

In particular, the HER2-positive subtype is an aggressive form of breast cancer and generally associated with poor prognosis[5]. Overexpression of the HER2 membrane protein occurs in 18-20% of breast tumors[6]. Assessment of the HER2 status is therefore important to determine whether a patient qualifies for the targeted antibody-based anti-HER2 therapy, an adjuvant to standard breast cancer chemotherapy. HER2 status assessment is not only important in breast cancer but there is even evidence that HER2 testing may be beneficial in gastric[7] and gynecological[8] malignancies. Techniques such as immunohistochemistry (IHC), polymerase chain reaction (PCR), enzyme-linked immunosorbent assay (ELISA), Southern blotting or *in situ* hybridization (ISH) are used to determine the HER2 status. As recommended by the American Society of Clinical Oncology / College of American Pathologists (ASCO/CAP),[9] IHC (protein level) and FISH (gene level) are the most common techniques for HER2 assessment in diagnostics.

IHC continues to be the 'gold standard' for HER2 testing in most diagnostic laboratories despite the bottlenecks outlined below. IHC relies on antibodies binding to the HER2 antigen on the cell membrane and their subsequent detection by a labelled secondary antibody. The samples are graded from 0 to +3 according to the staining intensity of the sample (0: no expression, +1: weak expression, +2: equivocal, and +3: high expression). Only samples scored with grade +3 qualify the patient for the targeted anti-HER2 therapy. However, approximately 20% of results obtained using IHC for HER2 testing are inaccurate[10]. IHC results are not quantitative, since the staining intensity is assessed by eye resulting in subjective grading of the tumor and thus the results often differ between different laboratories. Further, loss of sensitivity secondary to antigenic alterations[11] caused by standard fixation procedures and batch-to-batch variations of the antibody[12] can alter the staining performance. In addition, staining results are often not clear and reported as equivocal. In this situation, a second test – ISH as suggested by the ASCO/CAP guidelines[10] – is needed to assess the *HER2* amplification status, delaying diagnosis and requiring additional samples (Fig. 1b).

In contrast to IHC, *in situ* hybridization is a cytogenetic method and, specifically for HER2 testing, relies on ISH probes, labelled nucleic acid sequences, that hybridize to the locus of the *HER2* gene on chromosome 17 and the centromeric region of chromosome 17[13], Fig. 1c. It was reported that 90-95% of breast carcinomas that overexpress HER2 do so secondary to *HER2* gene amplification[14,15]. An ISH result is positive if HER2/Cen17 ratio ≥2 or the average *HER2* copy number ≥6 signals/cell, and negative if HER2/Cen17 ratio <2 according to the 2013 ASCO/CAP guidelines[10] (Fig. 1b). ISH is a specific method[16] and since each ISH signal represents a single copy of a gene, it is considered more quantitative and accurate than standard IHC. Despite these merits of ISH, it is not a routine diagnostic procedure; it is primarily used to assess the HER2 status of the patient samples, which were reported as IHC equivocal[10]. This is the case because ISH tests are time consuming ranging from one to several days, the reagents are expensive, especially the FISH probes, and highly trained personnel are required in the diagnostic laboratories.

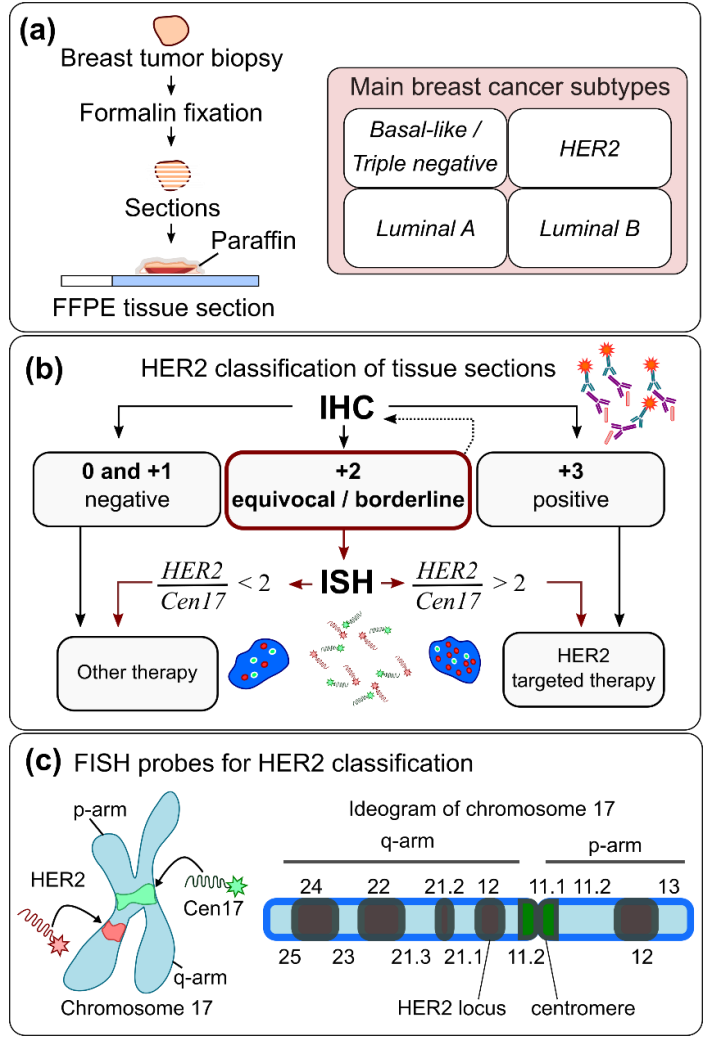

*FIG. 1. HER-2 classification of FFPE tumor samples. (a) Preparation steps for FFPE tissues from solid tumors. (b) Representative HER2 testing algorithm in diagnostics. (c) FISH probes for HER2 classification bind to the locus of HER2 gene (red) and centromeric region (green) of chromosome 17. The HER2 locus is located on the q12 arm of chromosome 17.*

With the objective of making ISH pervasive in diagnostic laboratories, the microfluidic community has been developing miniaturized implementations for ISH-based analysis of cells and tissue sections. For analysis of cancer cell lines, devices were presented to miniaturize the assay of both adherent[17–19] and non-adherent cells[20–22], while other implementations focused on automation of the assay[23,24].

For FISH-based microfluidic analysis of FFPE tissue sections only a few implementations have been reported; Kao *et al.* as an example focused on automation of the assay[25] and reduced the FISH probe consumption from 10 μL to 2 μL with 16 h static hybridization resulting in a turnaround assay time of 20 h. Using flow-based FISH probe incubation for 4 h, Nguyen *et al.*[26] reduced the FISH probe consumption to 2 μL and the turnaround assay time from >24 h to 8 h. They use a PDMS device, which is contacted with the tissue section and FISH probes are loaded into the microchannels for hybridization. In a recent study[27], they presented extra short incubation microfluidic assisted (ESIMA) FISH for HER2 testing of tissue sections with an incubation time of 35 min and probe consumption of 3 μL per test using the recently reported Instant Quality (IQ) FISH buffer[28].

While microfluidic implementations solved some bottlenecks of FISH by reducing the reagent consumption, the turnaround time or the work load of the FISH assay, those 'closed' microfluidic FISH methods are limited due to the need for a microfluidic chip that is clamped to the tissue section. Further, the geometry of the microchannels are fixed and therefore do not adapt to the varying histology between tissue sections from different sources.

Here, we present a specific class of microfluidic-based ISH methods for rapid, economic and efficient molecular subtyping of tumor sections using a microfluidic scanning probe (MFP). In contrast to the traditional microfluidic devices, the MFP is a scanning, non-contact technology, which allows reagents to be shaped in the 'open space'. This allows microfluidics-based FISH to be performed on cells in selected regions of the tissue section without physical contact between the device and the cytological sample. In previous work[29], we introduced the concept of microfluidics-based FISH (μFISH) with an MFP where we demonstrated spatially multiplex micro-scale FISH on cells. To demonstrate the diagnostic use of μFISH with an MFP, here we present methods and protocols for rapid HER2 status assessment of FFPE cell blocks and tissue sections while also using small volumes of FISH probes.

## II. MATERIAL AND METHODS

### A. Microfluidic probe platform

The platform comprised a standard inverted microscope, sub-micrometer precision stages, syringe pumps connected to a microfabricated silicon-glass hybrid MFP head, mounted to a custom-made *z*-stage (Fig. S1, supplementary material). The head used for all experiments was a microfabricated device that localizes liquids on the cytological samples. The head itself contained four microchannels, which open up at the apex of the head and form apertures: two for injection and two for aspiration of liquids to and from the surroundings. The channel dimensions of the inner two apertures were $100 \times 100$ μm$^2$ and the dimensions of the outer two apertures $100 \times 200$ μm$^2$. The microfabrication of the head was described elsewhere[30]. The head was connected to glass syringes (Hamilton, 1705 TLLX) via adapters and tubings (IDEX, Tygon) and an 8-port circular Dolomite connector. Via control of the linear stages (Lang GmbH, Hüttenberg, Germany), *x*- and *y*-position of the head relative to the substrate was controlled with sub-micrometer accuracy. The MFP head was mounted to a z-stage on the microscope via a mounting plate (Fig. S1, supplementary material). An environmental chamber (Life Imaging Services GmbH, 'The Cube and the Box') for temperature control was placed around the platform.

**B. Preparation of FFPE samples for FISH**

FFPE MCF-7 and BT-474 cell blocks (CellMax™ FFPE Control Cell Line Slides - MCF-7 and BT-474) were purchased from AMS Biotechnology and 5 μm thick breast tumor tissue microarrays (Her2B) were purchased from US Biomax Inc. A datasheet with HER2 IHC data was provided from the vendor. The baking and heating steps were performed on a hotplate (Cimarec$^+$™, Cat. No. BARNHP88857105). First, the FFPE sample was baked at 60 °C for 1 h. Next, the sample was dewaxed in xylene (Merck, Cat. No. 1082972500) for 10 min. Then the sample was transferred to an ethanol-containing (Sigma Aldrich, Cat. No. 02860) glass beaker for 3 min and then dried completely at RT. For antigen retrieval, the sample was transferred in a solution of 0.08× saline sodium citrate (SSC) in ddH$_2$O heated to 98 °C for 20 min (from 20× SSC stock, 3 M NaCl, 0.3 M sodium citrate, pH 7.0, Thermo Fisher Scientific, Cat. No. 15557-044). Then the sample was rinsed with ddH$_2$O for 5 min and dried at RT. Next a hydrophobic barrier was drawn around the section with a pap pen (Sigma, Cat. No. Z377821) and dried at RT. The section was then digested using pepsin at 37 °C for 10-20 min (Milan Analytica AG, Cat. No. PSS060) and subsequently rinsed with ddH$_2$O for 5 min. Subsequently, the sample was dried and the pre-treated samples were covered with FISH buffer (Sections III.A & C, KBI-FHB, Leica Biosystems) or an aqueous ethylene carbonate solution (Section III.B.), covered with a cover slip and denatured at 65-72 °C for 10 min. After this step, the cover slip was removed, the slide rinsed with prewarmed 1×SSC and hybridization experiments were performed.

For benchtop FISH control experiments, the FISH probes were diluted as specified by the manufacturer's protocol in FISH buffer (KBI-FHB). This hybridization mix was subsequently pipetted onto the pre-treated FFPE cell blocks, which were covered with a cover slip and denatured at 72°C for 10 min. After denaturation the slide was incubated fpr hybridization at 37°C overnight. After hybridization, the slides were washed 2 times to wash non-specifically bound probes with 0.1 % IGEPAL CA-630 in 2× SSC (v/v) for 1 min at RT and with 0.3 % IGEPAL CA-630 in 0.4 × SSC (v/v) at 72°C for 2 min (Sigma Aldrich, Cat. No. I8896). Subsequently, an additional wash was performed with 1× SSC at RT for 1 min. The cells were then mounted with mounting medium containing DAPI (Thermo Fisher Scientific, Cat. No. S36938) for inspection.

**C. FISH probe preparation and FISH probe loading into the MFP**

The Cen17 probes were prepared from a mix of seven centromere 17-specific oligonucleotides with 3' and 5' Cy3 labels (SI Table S2), which were purchased from Integrated DNA Technologies (Coralville, Iowa). The sequences were adapted from Ref.[31] The 10× FITC-labelled HER-2 SureFISH probe was purchased from DAKO (Cat. No. G100046G-8). For recirculation experiments SSC buffer was used as the hybridization buffer. For HER2 classification, an ethylene carbonate (EC) buffer based on Ref.[28] was prepared: 15% v/v ethylene carbonate (Sigma, Cat. No. E26258), 20% v/v dextran sulfate (Sigma, Cat. No. D8906), 0.67×SSC and 600 mM NaCl (Sigma, Cat. No. 71378) in nuclease-free water (Sigma, Cat. No. W4502).

For all experiments the Cen17 probes were used at a total concentration of 28 nM. The HER2 probe was diluted 1:10 in Cen17 containing EC-buffer. These FISH probes were denatured at 75 °C (formamide) or 66-68 °C (ethylene carbonate) for 5–10 min in a PCR cycler (QIAGEN, Rotor-Gene Q). To visualize the footprint indirectly, Hoechst dye was added to the probes at a concentration of 0.2 μg mL$^{-1}$ (Thermo Fisher Scientific, Cat. No. H3570). Subsequently these denatured probes were pipetted into a PCR cap and from there aspirated into the inner injection aperture of the MFP head.

**D. μFISH protocol**

One hour prior to performing the experiment, the environmental chamber around the MFP platform was heated to 45°C. After FISH probe loading into the MFP head, the pre-treated sample was transferred to the sample holder of the MFP platform, and the apex of the head was positioned 20-30 μm above the section. Subsequently, a hierarchical hydrodynamic flow confinement (hHFC) of FISH probes was contacted with the substrate as depicted in Fig. 2. In this flow configuration, two processing liquids were contacted with selected cells simultaneously: FISH probes were confined between the inner apertures and injected with flow rate $Q_{i2}$ and the wash buffer (1× SSC) was confined between the outer pair of apertures and injected with $Q_i$. For *HER2* assessment experiments, the flow rates were set to 10, 7 nL min$^{-1}$ ($Q_{i1}$, $Q_{i2}$) and –7, –100 nL min$^{-1}$ ($Q_{a1}$, $Q_{a2}$), accordingly. For the chromosomal enumeration and recirculation experiments, the flow rates were set to 100, 200 nL min$^{-1}$ ($Q_{i1}$, $Q_{i2}$) and –0.2, –2 μL min$^{-1}$ ($Q_{a1}$, $Q_{a2}$), respectively. After 5-15 min of interaction of the hHFC with the cells (equivalent to 5-15 min of incubation), $Q_{i2}$ was stopped, and the nuclei were washed with 1× SSC flowing between the outer apertures for 2 min. To avoid reflections, the head was positioned away from the slide before imaging. The footprint for all MFP-based FISH experiments was 0.096 mm$^2$.

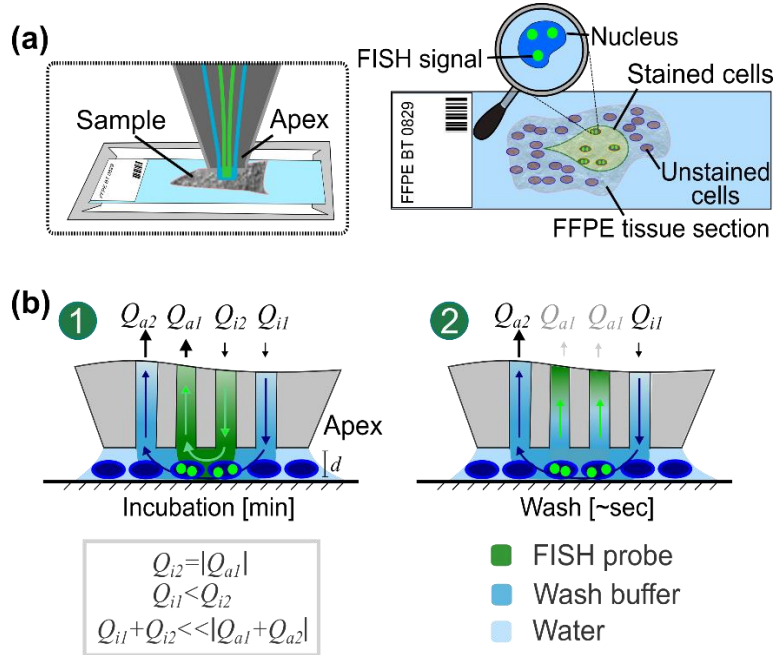

*FIG. 2. Schematics of the µFISH assay implemented with an MFP on a tissue section. (a) At the apex of the scanning head, FISH probes are confined hydrodynamically. FISH signals are present in cells in the region of the flow confinement (footprint). (b) Hydrodynamic flow confinements: In flow configuration 1, FISH probes (green) are confined between the inner pair of apertures. During incubation, the probe confinement is in contact with the cytological sample. After incubation, the inner flow is stopped and the cells from the sample are rinsed with the wash buffer (blue) from the outer injection aperture (flow configuration 2). Subsequently, FISH signals are inspected.*

**E. Image acquisition and processing**

Both, the endpoint observation for conventional FISH and the real-time observation for µFISH were performed using an inverted microscope at 10×, 40× and 60× magnification (Nikon Eclipse Ti-E with objectives CFI Plan Fluor DLL 10×, ELWD 40× and ELWD 60×, respectively). An LED lamp (Sola, Lumencor) and filter cubes FITC (F36-525), Cy3 (F36-542) and DAPI (F36-498) from AHF Analysentechnik were used for excitation and emission control. Image acquisition was performed using a Hamamatsu ORCA-flash 4.0 camera controlled with the NIS Elements Basic Research software (Nikon Instruments Europe, V.4.0). Image processing of raw images as well as merging were done using the open-source FIJI (ImageJ) software (http://fiji.sc/Fiji) (Fig. S3, supplementary material).

## III. RESULTS

**A. Oligo-based probes for chromosomal enumeration in FFPE tissue sections with <5 min incubation**

We aimed to enumerate the copy numbers of chromosome 17 in FFPE tissue sections, and for this we adapted the MFP-based FISH method[29]. In contrast to the 'fresh' cells we used for our previous study, FFPE samples undergo a series of preparation steps affecting the sample integrity[32–38]. Formaldehyde fixation of cytological samples is associated with DNA modifications[32], DNA fragmentation[33] and DNA-protein crosslinking[34], which affects the DNA integrity. Further, the crosslinked matrix can reduce the FISH probe mobility, i.e. the FISH probe penetration through the sample and the target accessibility, thereby lowering the efficiency in hybridization of probes to the target. To enable rapid FISH-based chromosomal enumeration on FFPE tissue sections, we: (i) adapted the FISH probe and moved to using short oligonucleotides, and (ii) optimized the composition of the hybridization buffer. DNA mobility in the cell and cell nucleus decreases as the length of the DNA increases[39]. Similarly, the mobility of FISH probes in the cell decreases with increasing length and thus it has been found that an optimal length for oligonucleotide FISH probes is 18-50 nucleotides (nt)[40]. Longer probes will result in increased hybridization times and low synthesis yields; on the other hand, shorter probes will result in reduced specificity[40]. To yield rapid FISH we therefore used 18 nt long oligonucleotide probes as the Cen17 probe (Table SI, S2 in supplementary material) to visualize the centromeric region of chromosome 17. Further, we prepared a formamide-free hybridization buffer and used the traditional FISH buffer (~50% formamide) only for denaturation of the chromosomal DNA of the sample. Interestingly, FISH signals appeared as early as ~30 s after initiating the flow of FISH probes over cells in the selected area of the tissue sections, in our experiments. Strong signals of the Cen17 probes in the nuclei of FFPE breast tumor tissue sections were detected within < 5 min (Fig. 3). Thus, the short DNA FISH probes are characterized with a high mobility through the crosslinked matrix. In addition, the hybridization process itself is likely accelerated by the change of hybridization buffer to the formamide-free composition. Formamide is a denaturation agent, which interferes with the hydrogen bonds forming during hybridization[41], thereby slowing down the hybridization.

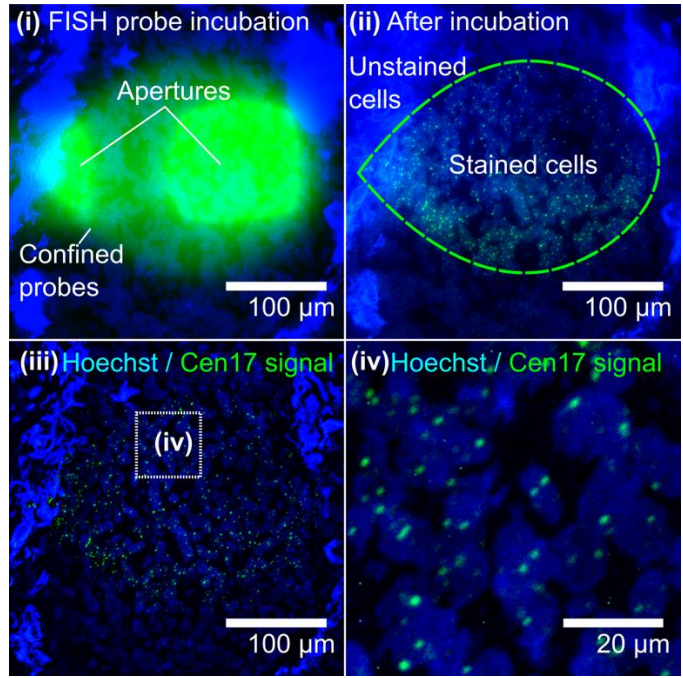

*FIG. 3. µFISH-based chromosomal enumeration in FFPE tumor sections (i) In situ hybridization occurs only in the cell nuclei (blue) within the area of the flow confinement (green). The tear shape is caused by the asymmetry between the injection and aspiration rate (i,ii). Fluorescent micrographs of Cen17 FISH signals (green) before (ii) and after image processing (iii,iv).*

**B. µFISH-based HER2 testing for diagnosis of breast cancer tissue sections.**

We then posed the question, whether this MFP-based method can be used in a diagnostic context, for HER2 testing on FFPE cell blocks and tissue sections. To achieve this, we used a mix of oligo-based HER2 and Cen17 probes. However, *HER2*-specific probes did not create FISH signals within 5 min incubation with MFP-based FISH on FFPE samples, when diluted in in 1× SSC. The chosen HER2 probes are commercially available synthetic oligonucleotide probes and specifically designed to the *HER2* locus[42]. Cen17 FISH probes bind repetitive elements of chromosome 17 and each probe sequence has ~100 targets per chromosome. In contrast, the *HER2* locus contains fewer repetitive elements and each *HER2*-specific probe sequence binds a unique target within the *HER2* locus. Thus, the probability of a Cen17 finding its target is higher than the probability of a HER2 probe finding its target. Further, HER2 probes used here are long oligonucleotide probes[42,43] with a length of ~ 150 nt. Since DNA mobility is reduced with increasing length[39], the HER2 probes require more time for hybridization than the 18 nt Cen17 probe.

To accelerate the hybridization reaction, we therefore modified the buffer by adding dextran sulfate as a volume exclusion agent[44] and ethylene carbonate (EC), adapted from the Instant Quality (IQ) FISH method[28]. Performing FISH probe incubation of the probe mix in EC-buffer using the MFP, we were able to assess the HER2 status of both FFPE cell blocks and breast cancer tissue sections within 10 min (cells) and 15 min (tissue) incubation time (Fig.

4.a,b). Even at flow rates of 7 nL min$^{-1}$ we were able to establish a stable flow confinement. Further, since the probe mix is viscous, the flow confinement was still in contact with the cells at these low flow rates and a surface-to-apex distance $d = 20$ µm. After less than one minute, the Cen17 signals were saturated and HER2 signals appeared within 15 min incubation (Fig. 4.a,b).

We specifically chose MCF-7 (*HER2* negative) and BT-474 (*HER2* positive) cell blocks for validation of the MFP-based FISH method and quantified the HER2/Cen17 ratios in 20 cells as recommended by the ASCO/CAP guidelines after both benchtop FISH using a pipette and µFISH experiments (Fig. S4, supplementary material for fluorescent micrographs after benchtop experiments). The HER2/Cen17 ratios calculated for both on-bench and MFP-based FISH were experiments were comparable for as illustrated in the graph in Fig. 4.c. The ratio was 0.82 (MFP) and 0.64 (benchtop) in MCF-7 cells and in BT-474 cells, the ratio was 2.8 (MFP) and 3.0 (benchtop).

In addition, we tested the µFISH method on FFPE tissue sections and assessed the HER2/Cen17 ratio in 20 cells from two tissue cores from a tissue microarray. As depicted in Fig. 4.d the HER2 status assessed with µFISH matched the IHC data of the datasheet from the tissue microarray.

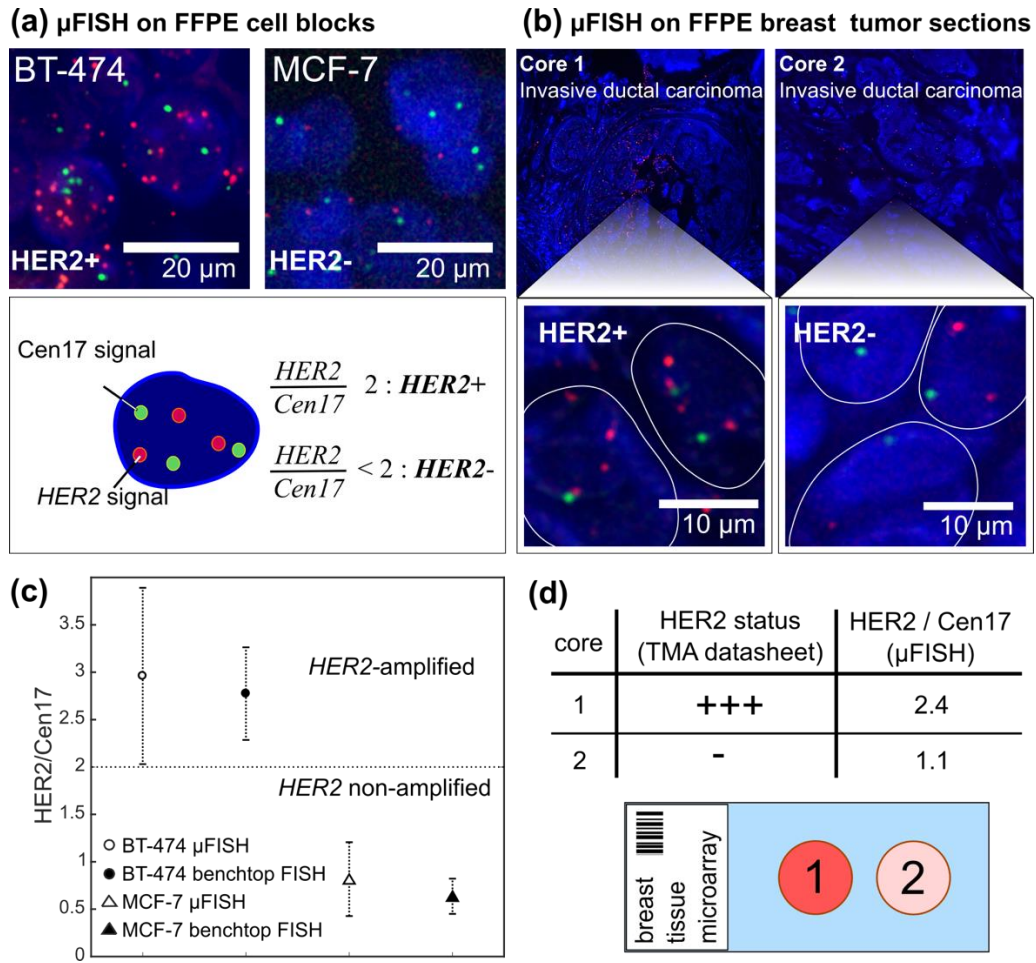

*FIG. 4. µFISH-based HER2 assessment in FFPE cell blocks and tumor sections. (a) Fluorescent micrographs for qualitative validation of µFISH for HER2 assessment on HER2 amplifying (BT-474) and HER2 non-amplifying (MCF-7) FFPE cell blocks. (b) Validation of µFISH-based HER2 assessment on nuclei (outlined) of HER2-amplified (left) and HER2 non-amplified (right) FFPE invasive ductal carcinoma tissue cores from a tissue microarray. (c) HER2/Cen17 ratios in MCF-7 and BT-474 cells in FFPE cell blocks assessed with benchtop and MFP-based FISH. Error bars present the standard error of the mean for n=20 measurements. (d) Quantified HER2/Cen17 ratio from 20 cells of the tissue cores illustrated in (b) comparing to the HER2 status from the tissue microarray datasheet.*

**C. FISH probe recirculation for efficient use of FISH probes**

For a more economic use of FISH probes, we tested the concept of recirculation on FFPE cell blocks. In a classical flow confinement, a processing liquid is confined between two apertures at the apex of the MFP head and the surface, in the presence of an immersion liquid. Since the aspiration rate is set to be approximately 3-fold higher than the injection rate to ensure a stable flow confinement, immersion liquid will be aspirated together with the processing liquid into the aspiration channel, resulting in an ~3-fold dilution of the processing liquid. The addition of two extra apertures, allows the processing liquid to be nested within a confinement of a second liquid, resulting in a hierarchical flow confinement[46]. Specifically, for MFP-based FISH, in a hierarchical hydrodynamic flow confinement the FISH probes are confined between the inner two apertures, while between the outer two apertures, a wash buffer is confined to nest the FISH probes (Fig. 5a). Since the outer aspiration rate is high with $Q_{a2}>>Q_{a1}$ and maintains a stable flow confinement, the inner injection and aperture rates can be set to $Q_{i2}=|Q_{a1}|$. Due to this symmetry between $Q_{i2}$ and $Q_{a1}$, a liquid injected with $Q_{i2}$ is aspirated by $Q_{a1}$ with minimal dilution. Therefore, a liquid confined between the inner two apertures can be re-injected and thus recirculated after inversion of the flow rates: $Q_{i2}\leftrightarrow Q_{a1}$, $Q_{i1}\leftrightarrow Q_{a2}$ (Fig. 5).

The graph in Fig. 5b illustrates the probe dilution per recirculation cycle for 1 µL of Cen17 probes in 1× SSC with a probe injection rate $Q_{i2}$ of 0.2 µL min$^{-1}$, 5 min switching intervals and a surface-to-apex distance $d = 20$ µm. We measured the fluorescence intensity for each recirculation cycle and normalized the intensity with respect to the initial fluorescence intensity of the confined probes. As illustrated in Fig. 5b, the FISH probes are diluted by ~ 6% per recirculation cycle. This dilution is likely due to the slight asymmetry of the flow profile resulting in a small flow from the immersion liquid into the inner aspiration aperture and the loss of FISH probes aspirated into the outer aspiration aperture.

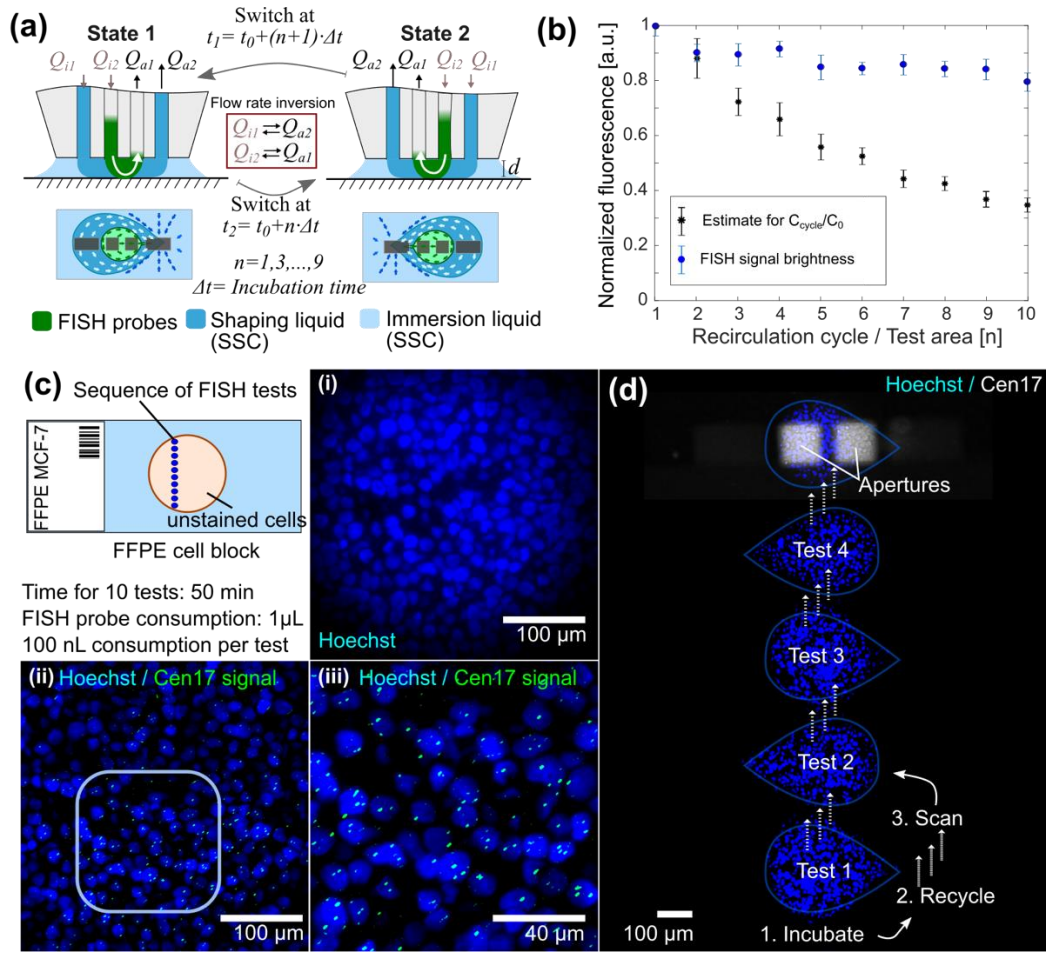

*FIG. 5. Recirculation of FISH probes using the MFP for FISH probe budgeting on MCF-7 cell blocks. (a) Switching between the two states for FISH probe (green) recirculation. (b) Dilution of FISH probes as a function of recirculation cycle for n=3 experiments (black dots) and averaged FISH signal intensity of n=12 FISH signals for 10 testing areas (blue dots). Error bars present the standard error of the mean. (c) Schematics of 10 FISH tests performed by recirculating 1 μL Cen17 probes on an MCF7 cell block. (i) Micrograph of nuclei within the last test region, test region 10, after 5 min incubation. (ii,iii) FISH signals in the processed nuclei in test region 10. (d) Micrograph of the MFP in operation.*

We next investigated the possibility of using a small volume of FISH probes for multiple FISH tests. In a proof-of-concept experiment we performed a sequence of ten FISH tests on an MCF-7 cell block by recirculating a defined volume of Cen17 probes ten times between the inner pair of apertures for 50 min with 5 min incubation time per FISH footprint (0.096 mm$^2$) (Fig. 5.c,d). At discrete time intervals of 5 min, we inverted the flow rates of FISH probes through the MFP microchannels and positioned the probe head to the next testing area. Cen17-specific signals were present in all cells within the 10 testing areas. Interestingly, even with 40% of the initial FISH probe concentration after 10 recirculation cycles, hybridization was efficient to yield FISH signals as strong as 80% of the initial signal intensity within 5 min incubation (Fig. 5.b,c).

## IV. DISCUSSION

Operating at flow rates of 7 nL min$^{-1}$ and 15 min incubation time results in a probe consumption of 105 nL per FISH test, which is a ~100-fold reduction compared with the standard FISH protocol. The assay turnaround time for MFP-based HER2 testing is <3 h, which is comparable with the turnaround time of an IHC assay. This potentially makes it feasible to perform both FISH and IHC-based tests simultaneously, rather than performing FISH tests only on equivocal patient's samples. Such a multimodal approach presents a more robust HER2 testing workflow than currently implemented in diagnostic laboratories and could have implications for not only breast cancer but also gastric[7] or gynecological[8] subtyping. This would further allow the healthcare professionals to make treatment decisions sooner, thereby saving precious time for the patient.

The processed area for MFP-based FISH measures 0.096 mm$^2$ and captures 300 cells from an FFPE tissue section depending on the amount of connective tissue, tissue type and cell density of the sample. For a FISH-based diagnosis, the 2013 ASCO/CAP guidelines suggest to at least count the FISH signals in 20 non-overlapping nuclei from the tumor cells[10]. Tumor cells in solid tumors are often clustered within the abnormal areas of the tissue and therefore in one MFP footprint, we capture enough cells for a diagnosis. A challenge of analyzing tissue sections is to perform the μFISH assay in the cancerous regions of the heterogeneous patient sample. However, abnormal regions of the tissue can be identified using the cell autofluorescence before MFP-based FISH.

In the MFP-based FISH implementation, convection-based delivery of reactants accelerates the reaction by enhancing the mass transport; the reagent consumption is, however, dependent on the reaction duration, because the incubation step is flow-based. To overcome this, we made use of the capability of the MFP to recirculate liquids and demonstrated that a finite volume of FISH probes can be recycled, and the probes are used more efficiently. Recirculation of probes has several advantages. It allows multiple areas of a heterogeneous sample or multiple samples to be probed using one volume of probes. Further, even slow-hybridizing FISH probes such as long FISH probes of > 500 nt can now be hybridized with the MFP without increasing the probe consumption.

## V. CONCLUDING REMARKS

We introduced methods and protocols for rapid detection of an important breast cancer biomarker, human epidermal growth factor receptor 2 (HER2), in tissue sections. We make use of a microfluidic probe technology and perform localized FISH probe incubation with cells in selected regions of tissue sections and cell blocks. This is done by confining FISH probes hydrodynamically on the tissue section at the micrometer length-scale without needing to construct physical walls. By using minimal probe amounts and also minimal amounts of the scarce cytological sample, the MFP-based FISH implementation presents a valuable method for diagnosis of breast cancer biopsies.

The focus of this work was to overcome the rate-limiting step in FISH, the probe hybridization, by using an MFP. However, combining recently reported features for liquid heating[47] and liquid switching[48] in the MFP head, it is possible to implement the entire FISH protocol and automate the assay with the MFP. A shortcoming of the used device is currently the low throughput. There is the possibility to form multiple confinements and process a larger number of samples simultaneously by adapting the design of the MFP device. The shape and the footprint dimensions can be adapted by changing the aperture geometries which could range from $1 \times 1$ µm$^2$ up to $1.2 \times 0.3$ mm$^2$ for the vertical MFP[49,50]. To further reduce the time of tests, one approach is to increase the concentration of the FISH probes. Phung *et al.* recently focused a high concentration plug of FISH probes with isotachophoresis in a microfluidic channel for FISH analysis of bacteria[51]. Combining this approach with the HFC to move a 'static' high concentration plug for more rapid FISH analysis might be possible by applying a counterflow.

Going forward we expect to adapt the method for multiplexed RNA FISH for research applications using branched RNA probes. The recently developed RNA ISH methods revealed new biomarkers, which may become diagnostic or prognostic markers in the future[52–54]. Overall, µFISH supports our vision of tissue microprocessing to implement several bio-analytical tests on limited biopsy samples.

**SUPPLEMENTARY MATERIAL**

See supplementary material for further information on the MFP platform, image processing with ImageJ, bench-top FISH controls and Cen17 probe sequences.

**ACKNOWLEDGMENTS**

This work was supported in part by the European Research Council (ERC) Starting Grant, under the 7th Framework Program (Project No. 311122, BioProbe). We would like to thank Xander van Kooten, David Taylor, Robert D. Lovchik, Yuksel Temiz and Ute Drechsler for help in experiments and device fabrication. We would also like to thank

Anne-Marie Cromack and Lena Voith von Voithenberg for valuable comments on the manuscript, and Emmanuel Delamarche, Walter Riess and Prof. Andrew deMello (ETH Zurich) for their continuous support.

**COMPETING INTEREST**

The authors declare no competing financial interest.